\documentclass[pre,twocolumn,showpacs,showkeys]{revtex4}
\usepackage[ansinew]{inputenc}
\usepackage{amsmath}
\usepackage{graphicx}
\usepackage{hyperref}
\usepackage{dcolumn}
\begin{document}
% *****************************************************************************
% *****************************************************************************
% *****************************************************************************
\newcommand{\phantomFrac}{\vphantom{\frac{}{}}}
\newcommand{\e}{\varepsilon}
\newcommand{\subst}{\left.\phantomFrac\right|}
\newcommand{\imag}{\Im m\;}
\newcommand{\real}{\Re e\;}
\newcommand{\order}{\mathcal{O}}
\newcommand{\Az}[1][n]{A^{(#1)}}
\newcommand{\Jz}[2][n]{J_{#2}^{(#1)}}
\newcommand{\Atr}[1][n]{A_\text{Tr}^{(#1)}}
\newcommand{\An}[1][n]{{A_\text{N}}^{(#1)}}
\newcommand{\Ai}[1][n]{{A_\text{I}}^{(#1)}}
\newcommand{\Ae}[1][n]{\varepsilon_{A}}
\newcommand{\Aa}[1][n]{\chi^{(#1)}}
\newcommand{\Fi}[1][n]{\ensuremath{{f_\text{I}}^{(#1)}}}
\newcommand{\Fn}[1][n]{\ensuremath{{f_\text{N}}^{(#1)}}}
% *****************************************************************************
% *****************************************************************************
% *****************************************************************************
\title{Terahertz Bloch oscillator with suppressed electric domains: Effect of elastic scattering}
\author{Timo Hyart}
\email{thyart@student.oulu.fi}
\affiliation{Department of Physical Sciences, P.O. Box 3000,
  University of Oulu FI-90014, Finland}
\author{Kirill N. Alekseev}
\email{Kirill.Alekseev@oulu.fi}
\affiliation{Department of Physical Sciences, P.O. Box 3000,
  University of Oulu FI-90014, Finland}
\author{Ahti Lepp\"{a}nen}
\affiliation{Department of Physical Sciences, P.O. Box 3000,
  University of Oulu FI-90014, Finland}
\author{Erkki V. Thuneberg}
\affiliation{Department of Physical Sciences, P.O. Box 3000,
  University of Oulu FI-90014, Finland}
% -----------------------------------------------------------------------------
\begin{abstract}
We theoretically consider the amplification of THz radiation in a
superlattice Bloch oscillator. The main dilemma in the realization
of THz Bloch oscillator is finding operational conditions which
allow simultaneously to achieve gain at THz frequencies and to avoid
destructive space-charge instabilities. A possible solution to this
dilemma is the extended Limited Space-Charge Accumulation scheme of
Kroemer (H. Kroemer, cond-mat/0009311). Within the semiclassical
miniband transport approach we extend its range of applicability by
considering a difference in the relaxation times for electron
velocity and electron energy. The kinetics of electrons and fields
establishing a stationary signal in the oscillator is also
discussed.
\end{abstract}


\maketitle
% *****************************************************************************
% *****************************************************************************

\section{Introduction}

Terahertz radiation ($0.3-10$ THz) has enormous promising
applications in very different areas of science and technology such
as space astronomy, wideband communications and biosecurity, to name
a few (for recent reviews, see \cite{thz-reviews}). One of the main
challenges is to construct a coherent miniature source of THz
radiation that can operate at room temperature. Currently many
groups worldwide are working in this direction. Along with many
experimental and technological problems there are still several
fundamental problems to be solved. In particular, the traditional
lasing scheme is based on a population inversion between different
energy levels. A great recent achievement was the development of
quantum cascade lasers that can operate at THz frequencies employing
quantum transitions between the energy levels in multiple quantum
well heterostructures \cite{nature}. These quantum nanodevices
require a low temperature to achieve a significant population
difference. Continuous improvements in the design of THz quantum
cascade lasers allow to increase the temperature of operation above
$100$ K \cite{physicae}. However, it would be a quite difficult, if
ever possible, to reach population inversion at room temperature.
Really, the spacing of the energy levels, which is necessary to emit
radiation at frequencies of the order of $1$ THz, is comparable with
the room temperature in proper units. This simple observation makes
it attractive to study other suggestions for nano-devices that do
not require population inversion for the generation of THz
radiation.
\par
The Bloch oscillator is an inversionless THz laser that is based on
Bloch oscillations of miniband electrons in a dc-biased
semiconductor superlattice. The continuous operation of a Bloch
oscillator has still not been demonstrated in any experiments. The
development of a superlattice Bloch oscillator has been initiated by
suggestion of Esaki-Tsu \cite{Esaki} and theoretical analysis of
Ktitorov, Simin and Sindalovskii \cite{KSS}. The main obstacle in
the experimental realization is the formation of high-field electric
domains in superlattice.
\par
Nowadays the problem of THz Bloch oscillator again attracts much
attention (for review, see
\cite{wackerrew,knaKarlsruhe,knaThisMeeting}). There are several
interesting suggestions for modifications of the original scheme of
Bloch oscillator, which in principle should allow to reach THz gain
in superlattice without formation of destructive electric domains.
\par
In this report, we first briefly review the static electric
properties of superlattices and their influence on both small-signal
gain and electric instability in section~\ref{Sec_static}. In the
original part of this report we develop the idea of a
large-amplitude THz Bloch oscillator suggested by Kroemer
\cite{Kroemer}. We consider a dc-biased superlattice subjected to a
monochromatic ac probe field, so that the total electric field
acting on the miniband electrons is
%-------------Eq 1----------------------
\begin{equation}
\label{E-mono} E=E_{dc}+E_{ac} \cos{\omega t}.
\end{equation}
%--------------------------------------
Note that in a real device the probe field with the amplitude
$E_{ac}$ should be a mode of resonator tuned to a desirable THz
frequency. It is well known that while a superlattice device should
allow strong gain for a small ac field ($E_{ac}\rightarrow 0$) in
the wide range of frequencies from zero up to several THz,
simultaneously destructive electric instabilities would arise inside
superlattice (see section~\ref{Sec_static}). As Kroemer has
demonstrated for particular values of superlattice parameters, the
ac field with a large enough $E_{ac}$ can suppress domains but still
preserve significant THz gain \cite{Kroemer}. In ref. \cite{Kroemer}
the single scattering constant approximation was used. However, a
more realistic approximation allow two different relaxation
constants, $\gamma_v$ for the electron velocity and
$\gamma_\varepsilon$ for the electron energy.
\par
Using the technique of superlattice balance equations
(section~\ref{Sec_balance}), we re-examine the THz gain and the
criterion of electric stability for the case of large $E_{ac}$. Our
main aim is to find how robust the Kroemer scheme of Bloch
oscillator is against the effect of different relaxation constants
(section~\ref{Sec_main}). Here our main finding is that this effect
does not dramatically change the results obtained within the single
scattering constant approximation, if ratio
$\gamma_\varepsilon/\gamma_v>0.1$. Moreover, for
$\gamma_\varepsilon/\gamma_v\geq 0.5$ the differences from the
results obtained within the single relaxation time approximation are
insignificant.
\par
In section~\ref{Sec_time-scales} we consider the characteristic time
scales for the development of space-charge instabilities  and for
the growth of the ac field due to high-frequency gain in
superlattice. We come to the conclusion that it really is possible
to suppress a destructive accumulation of charges in the case of THz
oscillations with large amplitudes. However, our estimates still
demonstrate that it is very difficult to switch the device in the
regime of large amplitudes before electric domains would be formed.
Discussion devoted to this and other remaining problems in the
realization of THz Bloch oscillator is presented in the final
section.

\section{\label{Sec_static} Static electric characteristic and small-signal gain}
Let us first review the static electric properties of superlattices
using a single relaxation time. Nonlinear electron transport in a
superlattice with applied dc bias $E_{dc}$ is a well-studied
problem. The dc current through the superlattice depends on the dc
bias as \cite{Esaki}
%----------Eq 2-------------------------------------
\begin{equation}
I_{ET}=I_{\rm{peak}} \frac{2 E_{dc}/E_{cr}}{1+(E_{dc}/E_{cr})^2}.
\label{eq:ET}
\end{equation}
%-----------------------------------------------
Here $E_{cr}=\hbar/(ed\tau)$ is the Esaki-Tsu critical field,
%----------Eq no number-------------------------------------
$$
I_{\rm{peak}}=\frac{e n v_0}{2}
\frac{I_1(\frac{\Delta}{2k_bT})}{I_0(\frac{\Delta}{2k_bT})} A
$$
%-----------------------------------------------
is the peak current corresponding to $E_{dc}=E_{cr}$, $\tau$ is the
scattering time, $n$ is the density of carriers, $v_0=\Delta d/(2
\hbar)$ is the maximal electron velocity in the first miniband,
$\Delta$ is the miniband width, $d$ is the period of superlattice,
$T$ is the temperature, $A$ is the cross sectional area of
superlattice and $I_1(x)$ and $I_0(x)$ are the modified Bessel
functions.
\par
\begin{figure}
  \includegraphics[scale=0.45]{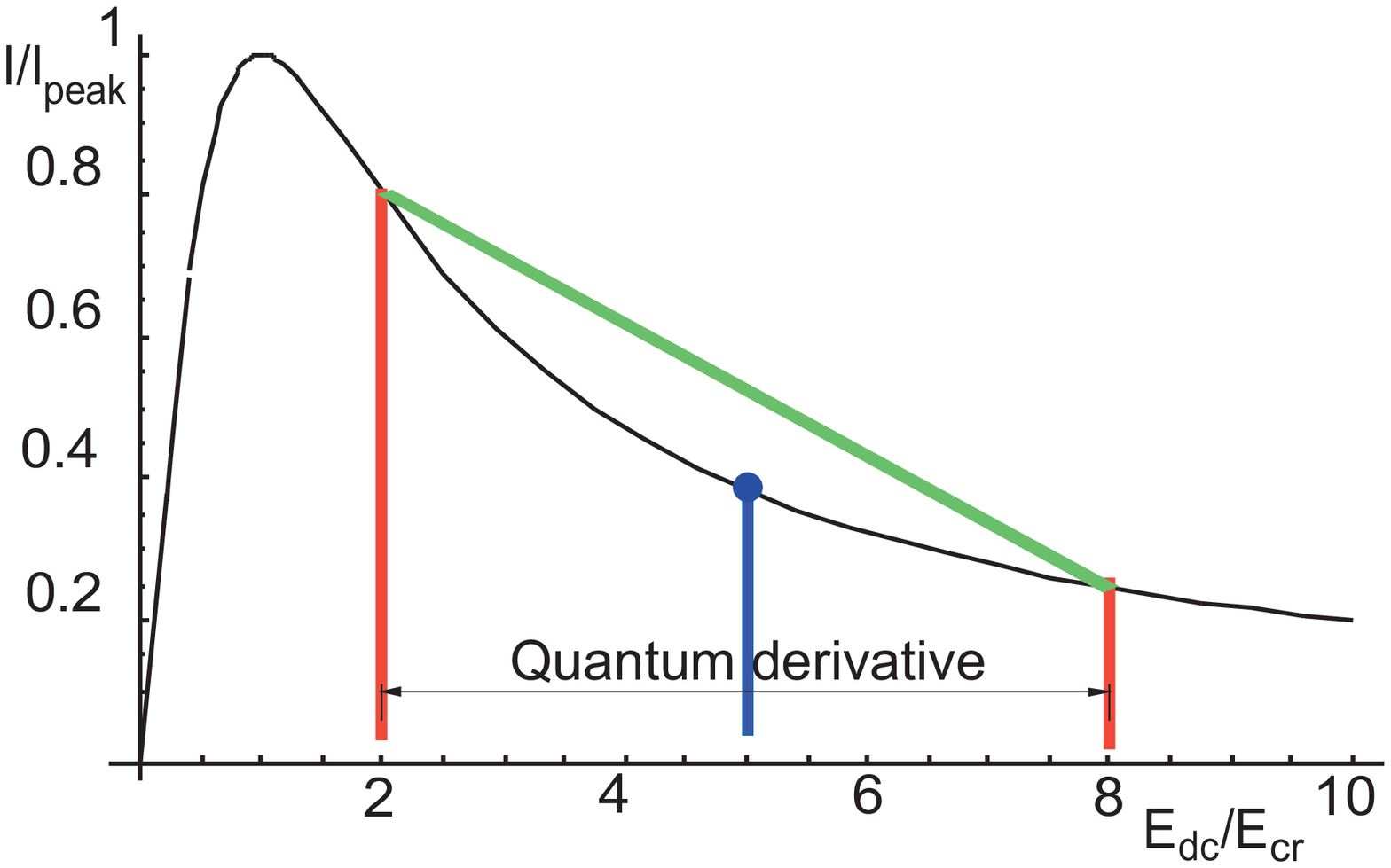}
  \caption{\label{fig:ET}(Color online) Esaki-Tsu current-field characteristic.
The figure also demonstrates the quantum derivative for the
frequency $\omega=3{\tau}^{-1}$ taken at the operation point
$E_{dc}=5E_{cr}$ (blue). Absorption at frequency $\omega$ is
proportional to the difference quotient, Eq. (\ref{quant-der1}),
which equals the slope of the inclined line (green).}
\end{figure}

The dependence of $I_{ET}$ on $E_{dc}$ is shown in
Fig.~\ref{fig:ET}. For $E_{dc}>E_{cr}$ the current-field
characteristic demonstrates negative differential conductance (NDC).
Note that $E_{dc}/E_{cr}=\omega_B\tau$ with
$\omega_B=edE_{dc}/\hbar$ being the Bloch frequency. Therefore the
condition of static NDC is also $\omega_B\tau>1$. For typical
semiconductor superlattices and for the applied dc bias $\simeq 1$
kV/cm, the Bloch frequencies belong to THz range \cite{Esaki}.
\par

In the following it is useful to consider the Esaki-Tsu current
(\ref{eq:ET}) as a function of voltage drop over one superlattice
period
%----------Eq 3-------------------------------------
\begin{equation}
\label{replica} I_{ET} (e E_{dc} d)=I_{\rm{peak}} \frac{2 e E_{dc}
d/\Gamma}{1+(e E_{dc} d/\Gamma)^2}
\end{equation}
%-----------------------------------------------
with $\Gamma=\hbar/\tau$.
\par
We turn to the consideration of electron transport under the action
of combination of dc and ac fields. The time-dependent current
induced in superlattice by the field (\ref{E-mono}) can be
represented as
%--------Eq 4------------------------------------------------
\begin{equation}
\label{current} I(t) =I_0^{\omega}+\sum_{h=1}^{\infty}[I_h^{\omega,
\cos} \cos(h \omega t)+I_h^{\omega, \sin} \sin(h \omega t)].
\end{equation}
%-----------------------------------------------------------------
Absorption of the probe ac field is proportional to the first
Fourier component of the time-dependent current $I_1^{\cos}$.
Negative absorption or gain corresponds to $I_1^{\cos}<0$.
\par
We start the discussion of small-signal gain ($E_{ac}\rightarrow 0$)
with the case of quasistatic interaction $\omega\tau\ll 1$. In this
case, it is easy to show that $I_1^{\cos}$ is proportional to the
slope of static VI characteristic \cite{wackerrew}. Therefore, a
choice of working point in the NDC portion of superlattice VI
characteristic results in small-signal gain for a quasistatic ac
probe field. Obviously, in the quasistatic case the gain does not
depend on the frequency. For typical superlattices at room
temperature $\tau\simeq 100$ fs. Therefore the quasistatic
approximation is valid for microwave fields.
\par
\begin{figure}
  \includegraphics[scale=0.45]{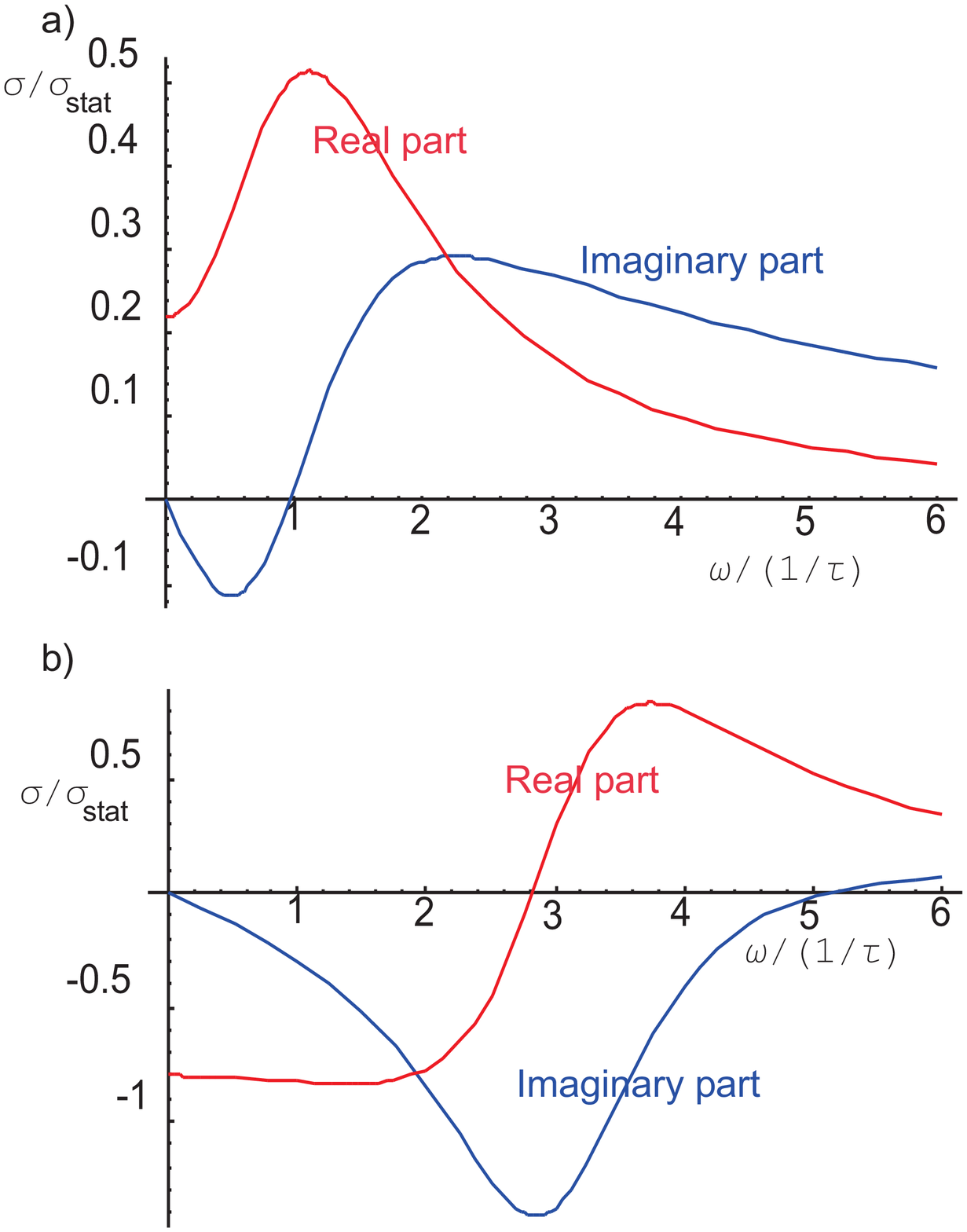}
  \caption{\label{fig:KSSkuva}(Color online) Real (red) and imaginary (blue) parts of
conductivity for (a) $E_{dc}=0.8 E_{cr}$  and (b) $E_{dc}=3E_{cr}$.
Note that for $E_{dc}=3 E_{cr}$ gain is possible almost up to
$\omega=\omega_B=3{\tau}^{-1}$.}
\end{figure}

For a higher frequency ($\omega\tau\gtrsim 1$), it is still possible
to have small-signal gain, but the magnitude of the gain becomes
frequency-dependent. Small-signal absorption is also often defined
using the frequency-dependent complex conductivity $\sigma(\omega)$:
$I_1^{\cos} \cos(\omega t)+I_1^{\sin} \sin(\omega
t)=\real\{\sigma(\omega) E_{ac} e^{-i \omega t} A\}$. It is easy to
see that absorption is proportional to $\real\sigma(\omega)$.
Complex high-frequency conductivity has been first calculated by
Ktitorov et al.\ \cite{KSS}
%----------Eq 5-------------------------------------
\begin{equation}
\label{eq:KSSres} \sigma(\omega)=\frac{1-i\omega \tau -
(E_{dc}/E_{cr})^2}{(1-i\omega\tau)^2+(E_{dc}/E_{cr})^2}
\sigma_{\rm{stat}},
\end{equation}
%-----------------------------------------------
where $\sigma_{stat}=I_{ET}/(A E_{dc})$ is the static conductivity
of the superlattice. Real and imaginary parts of the conductivity
(\ref{eq:KSSres}) are shown in Fig.~\ref{fig:KSSkuva}. Importantly,
for $E_{dc}\gg E_{cr}$ gain exists up to frequencies around the
Bloch frequency, with even a resonance near $\omega_B$. The physical
mechanisms of this high-frequency gain has been studied both in the
semiclassical picture with the help of electron bunches
\cite{Kroemer2} and in the quantum mechanical Wannier-Stark picture
\cite{willenberi}.
\par
Alternatively, the small-signal absorption can be calculated by
taking the so-called quantum derivative of the current-field
characteristic \cite{ignatov_q-deriv,wackerrew}  (Fig.~\ref{fig:ET})
%----------Eq 6-------------------------------------
\begin{equation}
\label{quant-der1} I_1^{\cos} = \frac{I_{ET}(e E_{dc} d+\hbar
\omega)-I_{ET}(e E_{dc} d-\hbar \omega)}{2 \hbar \omega} e E_{ac}d.
\end{equation}
%-----------------------------------------------
For $\omega\tau\rightarrow 0$ the difference quotient in the quantum
derivative becomes the usual derivative, and the result that small
frequency absorption is proportional to $dI_{ET}/dE_{dc}$ is
rediscovered. One immediate consequence of the quantum derivative is
that in order to get high-frequency gain we must have static NDC
(see Fig.~\ref{fig:ET}).
% _____________________________________________________________________________
% figure1
% _____________________________________________________________________________
%
% _____________________________________________________________________________
% figure2
% _____________________________________________________________________________
%
\par
Superlattices with static NDC, similarly as Gunn diodes, are
unstable against spatial space-charge fluctuations, which result in
formation of high-field electric domains. The possibility of
space-charge instability in the superlattice Bloch oscillator was
first pointed out by Ktitorov et al.\ themselves \cite{KSS}. Later,
it has been established within a simplified model by B\"{u}ttiker
and Thomas \cite{buettiker77} and in most consistent form by Ignatov
and Shashkin \cite{ignatov87}.
\par
The electric domains are believed to be destructive for
high-frequency gain in superlattices. Nevertheless, in series of
very interesting experiments French groups did observe gain at
reflection of microwaves from long superlattices with $n\simeq
10^{16}$ cm$^{-3}$ in a wide range of frequencies up to several
hundreds of GHz \cite{mm-gain-exp}. It seems that physical reasons
for these very interesting experimental observations are still not
well understood in the semiconductor community.
\par
On the other hand, for THz waves, Santa-Barbara group recently
reported a decrease of absorption, but still not gain, in an array
of short superlattices \cite{savvidis}. Experimental techniques for
monitoring the domains in superlattices are under development
\cite{lisauskas}.
\par
Now we turn to consider the regime of large probe field, where
high-frequency gain is not necessarily connected to the presence of
static NDC.

\section{\label{Sec_balance} Superlattice balance equations}

Until now we have discussed electron  transport in superlattices
assuming the existence of a single relaxation time. However,
experiments demonstrate that models employing two relaxation times
are more realistic. By using the tight-binding approximation and the
Boltzmann equation with two scattering times, the following balance
equations can be derived \cite{wackerrew}
%----Eqs 7----------------------------
\begin{eqnarray}
\frac{d}{dt}j(t)+\frac{d^2 e^2E(t)}{\hbar^2}
\varepsilon(t)&=&-\gamma_v j(t)
\nonumber\\
\frac{d}{dt}\varepsilon(t)- E(t) j(t) &=& -\gamma_\varepsilon
\big[\varepsilon(t)-\varepsilon_{eq}\big]. \label{balance}
\end{eqnarray}
%---------------------------------------
Here $\gamma_\varepsilon$ and
$\gamma_v=\gamma_\varepsilon+\gamma_{el}$ are the phenomenological
scattering constants for electron energy and miniband electron
velocity respectively, $\gamma_{el}$ is the scattering constant
describing elastic scattering events, $j(t)$ is the current density
and $\varepsilon(t)$ is the total miniband energy density of
electrons. Electrons are staying at the bottom of miniband for
$\varepsilon=-n \Delta/2$ and the upper edge is reached if
$\varepsilon=+n \Delta/2$, where $n$ is the density of electrons in
the first miniband. Average electron energy in thermal equilibrium
$\varepsilon_{eq}$ depends on the temperature, superlattice
parameters and carrier density.

The superlattice balance equations have been first introduced by
Ignatov and Romanov \cite{ignatov76}. Importantly,  equations
(\ref{balance}) can describe both transient ($t<\gamma_v^{-1}$) and
stationary ($t\gg\gamma_v^{-1}$) dynamics of miniband electrons
under the action of strong electric fields (The electric field is
turned on at $t=0$.). Earlier these equations have been employed in
the studies of decaying coherent Bloch oscillations
\cite{ignatov91}, absolute negative conductance \cite{ignatov82} and
efficiency of THz Bloch oscillator \cite{ignatov93}, as well as in
the analysis of such strongly nonlinear phenomena like
symmetry-breaking and chaos \cite{sb-chaos}.
\par

The solution of equations (\ref{balance}) for static field is a
scaled Esaki-Tsu characteristic. Its critical field is
\begin{equation}
E_{cr}=\hbar/(ed \tau_{\rm{eff}}),
\end{equation}
where $\tau_{\rm{eff}}=(\sqrt{\gamma_m \gamma_e} \ )^{-1}$. This new
definition of critical field is used in the following and it reduces
to the earlier definition, if $\gamma_v=\gamma_\varepsilon=1/\tau$.

In addition to parameter $\tau_{\rm{eff}}$, a parameter
$\nu=\gamma_\varepsilon/\gamma_v$ is needed to describe the system.
If $\nu=1$, there is no elastic scattering i.e. $\gamma_{el}=0$.

\par

Assuming strong electric field of the form (\ref{E-mono}), the
equations (\ref{balance}) have time-dependent stationary solution in
analytic form if $\gamma_v=\gamma_\varepsilon=1/\tau$. In this case
we find for the Fourier harmonics (\ref{current}) of the current
$I(t)=j(t) A$
%----------Eqs 8 -------------------------------------
\begin{eqnarray}
\label{anal_solu}
I_0^{\omega} &=& \sum_l J_l^2(\alpha) I_{ET}(eE_{dc}d+l \hbar \omega), \nonumber \\
I_h^{\omega, \cos} &=& \sum_l J_l(\alpha)\big[J_{l+h}(\alpha)
\nonumber
\\&&\hspace{13mm}+J_{l-h}(\alpha)\big]
I_{ET}(eE_{dc}d+l \hbar \omega), \nonumber \\
I_h^{\omega, \sin} &=& \sum_l
J_l(\alpha)\big[J_{l+h}(\alpha)-J_{l-h}(\alpha)\big] K(eE_{dc}d+l
\hbar
\omega), \nonumber \\
\end{eqnarray}
%-----------------------------------------------
where $J_n(x)$ are the Bessel functions, the summation is from
$-\infty$ to $+\infty$, $\alpha=eE_{ac}d/(\hbar\omega)$, $I_{ET}$ is
given by equation (\ref{replica}) and
%------Eq no number-------------------------------
$$
K(eEd)=I_{\rm{peak}} \frac{1}{1+(eEd/\Gamma)^2}.
$$
%-----------------------------------------------
\par
The solutions of the balance equations (\ref{anal_solu}) are the
same as the corresponding expressions for current components found
with the help of an exact formal solution of Boltzmann transport
equation with a single relaxation time \cite{chambers}. Moreover, it
is easy to see that for small ac field $E_{ac}$ (i.e. for $\alpha\ll
1$), $I_1^{\omega, \cos}$ becomes the quantum derivative defined by
eq.\ (\ref{quant-der1}).
\par
Finally, it is useful to find the expression for
the dc differential conductivity
%----------Eq 9-------------------------------------
\begin{eqnarray}
\frac{dI_0^\omega}{dE_{dc}}&=&\sum_l J_l^2(\alpha) \sigma_0(e E_{dc}
d+l \hbar \omega)\nonumber \\&=&\frac{2 I_{\rm{peak}}}{E_{cr}}
\sum_l J_l^2(\alpha) \frac{1-(e E_{dc} d+l \hbar
\omega)^2/\Gamma^2}{\big[1+(e E_{dc} d+l
\hbar \omega)^2/\Gamma^2 \big]^2},\nonumber\\
\label{diff-cond}
\end{eqnarray}
%-----------------------------------------------
where $\sigma_0 (e E_{dc} d)$ is the dc differential conductivity
taken from the Esaki-Tsu characteristic.
\par
Formulas (\ref{anal_solu}) and (\ref{diff-cond}) will be used in the
next section in the stability analysis of large-amplitude
oscillations in the Bloch oscillator for the case
$\gamma_v=\gamma_\varepsilon$. They can also be used to check the
numerical solution of the balance equations in the limiting case
$\nu=1$.

\par
Note that since in the case of static electric field the solution of
equations (\ref{balance}) was the scaled Esaki-Tsu VI characteristic
for all $\nu$, different scattering rates do not give any
qualitative changes to low frequency behavior. This is, however, not
necessarily the case for higher frequencies. The quantum derivative
result is valid only if $\nu=1$, although it works as a good
approximation, when $\nu \approx 1$: Qualitative changes in the high
frequency behavior can occur even for small signals, when $\nu$ is
significantly smaller than $1$. The changes in large amplitude
oscillation dynamics are even more interesting, because the
calculations for $\nu=1$ indicate possibility for gain and domain
suppression \cite{Kroemer}.

\section{\label{Sec_main} Large-signal gain and stability}

We solved the balance equations (\ref{balance}) numerically, found
the stationary time-dependent current (\ref{current}), and then
calculated the dc current component $I_0^\omega$, the absorption
$I_1^{\omega,\cos}$ and the dc differential conductivity
%---------Eq no number ---------------------------
$$
\sigma_0^\omega=\frac{dI_0^\omega}{dE_{dc}}.
$$
%----------------------------------------------
For $\nu=1$ these functions were found using analytic formulas
(\ref{anal_solu}) and (\ref{diff-cond}). In this case, we found
excellent agreement between these two approaches, as well as with
the results of reference \cite{Kroemer}.
\par
\begin{figure}
  \includegraphics[scale=0.51]{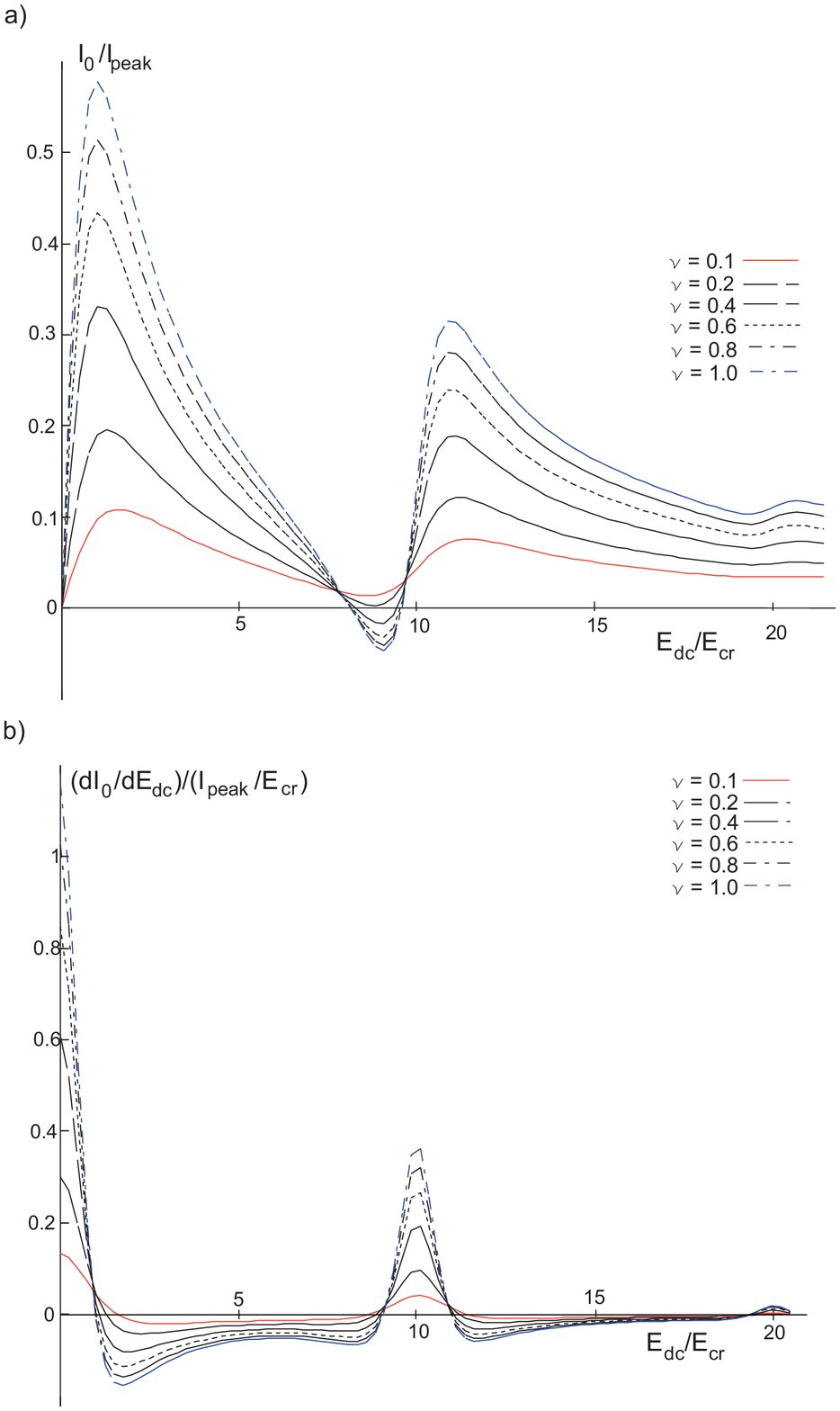}
  \caption{\label{fig:Inollavsele}
Typical current-field characteristic ($\omega\tau_{\rm{eff}}=10$,
$E_{ac}=10E_{cr}$) demonstrating local structures (a) with positive
differential conductivity (b) for
$\omega\tau_{\rm{eff}}-1<\omega_B\tau_{\rm{eff}}<\omega
\tau_{\rm{eff}}+1$. (Note that $\omega_B
\tau_{\rm{eff}}=E_{dc}/E_{cr}$.) The local structures become weaker,
when the ratio of scattering times decreases, but PDC can exist even
in larger range of parameters.}
\end{figure}

According to the formula (\ref{anal_solu}), the dc current can be
calculated with the help of photon replicas of the Esaki-Tsu
characteristic, and when the amplitude of the ac field increases,
the so-called Shapiro-like steps occur in VI characteristic
[Fig.~\ref{fig:Inollavsele} ($\nu=1)$]. Within these step-like
structures, regions of positive differential conductivity (PDC)
exists [Fig.~\ref{fig:Inollavsele} (b)]. When $\nu$ decreases, the
structures become less pronounced but the region of PDC becomes even
larger.
\par
The PDC should be considered as one of the conditions for electric
stability of the system \cite{Kroemer}. This is a sort of extension
of the Limited Space Charge Accumulation (LSA) mode of Copeland
\cite{copeland}, well-known in physics of Gunn diodes, to the case
of superlattices and THz frequencies.
% _____________________________________________________________________________
% figure3
% _____________________________________________________________________________
%
% _____________________________________________________________________________
% figure4
\begin{figure}
  \includegraphics[scale=0.57]{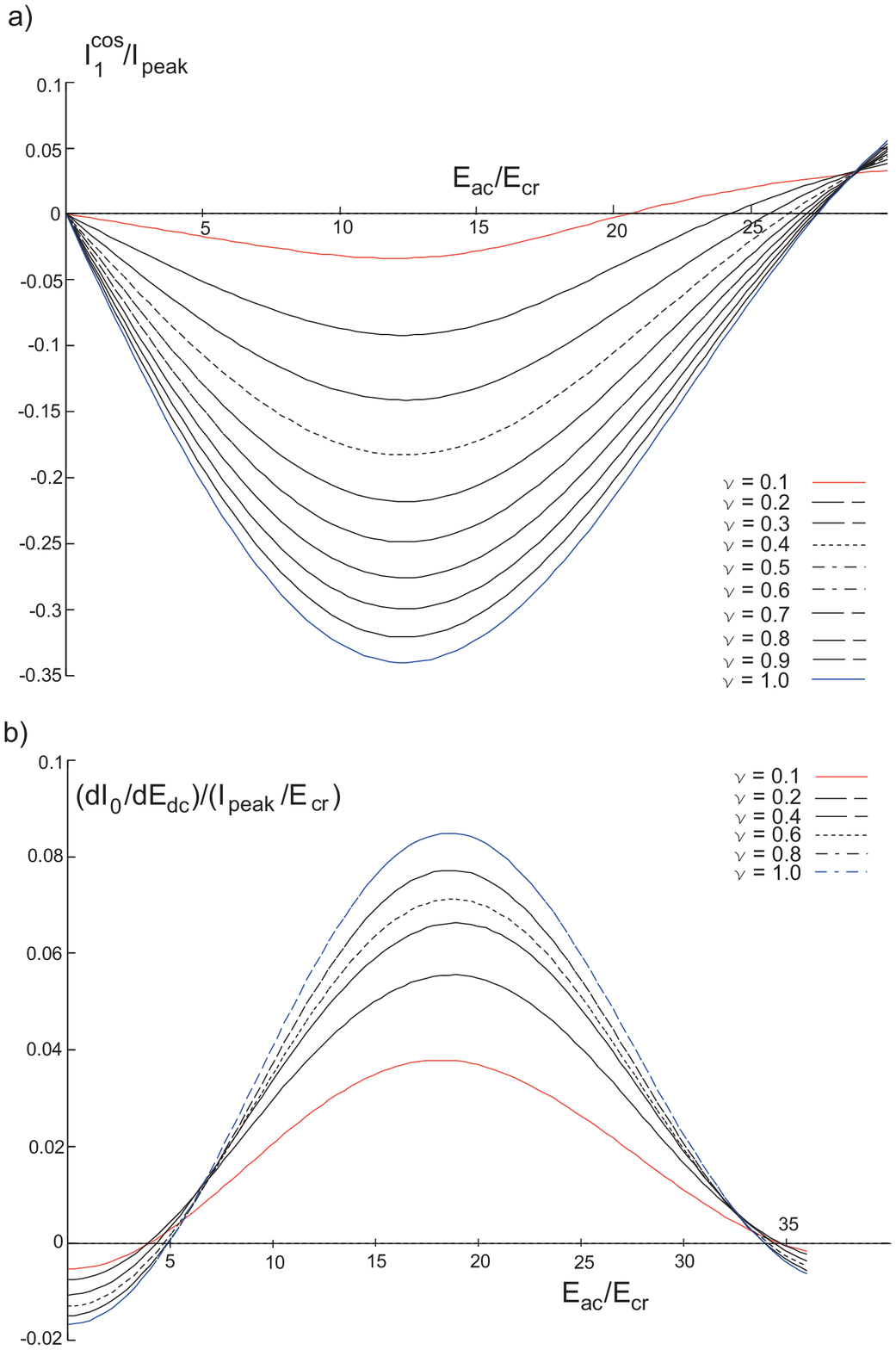}
  \caption{\label{fig:acnfunktiona}
Local PDC and gain can exist simultaneously in a wide range of
amplitudes $E_{ac}$. Dependencies of $I_1^{\omega,\cos}$ (a) and
$\sigma_0^\omega$ (b) on probe amplitude $E_{ac}$ are shown, when
$\omega \tau_{\rm{eff}}=10$ and $\omega_B \tau_{\rm{eff}}=\omega
\tau_{\rm{eff}}+0.8$. If $\nu=1$, dc differential conductivity is
positive for $5 E_{cr} \leq E_{ac} \leq 34 E_{cr}$ and absorption is
negative for $0 \leq E_{ac} \leq 27 E_{cr}$. This means that gain in
conditions of electric stability exists for $5\leq E_{ac}\leq 27
E_{cr}$. If $\nu$ is decreased, the local structures become weaker
and the magnitude of the absorption decreases: The gain in
conditions of electric stability occurs in a smaller range of
$E_{ac}$ values. The changes are small, if $\nu>0.5$.}
\end{figure}
% _____________________________________________________________________________
%
% _____________________________________________________________________________
% figure5
\begin{figure}
  \includegraphics[scale=0.44]{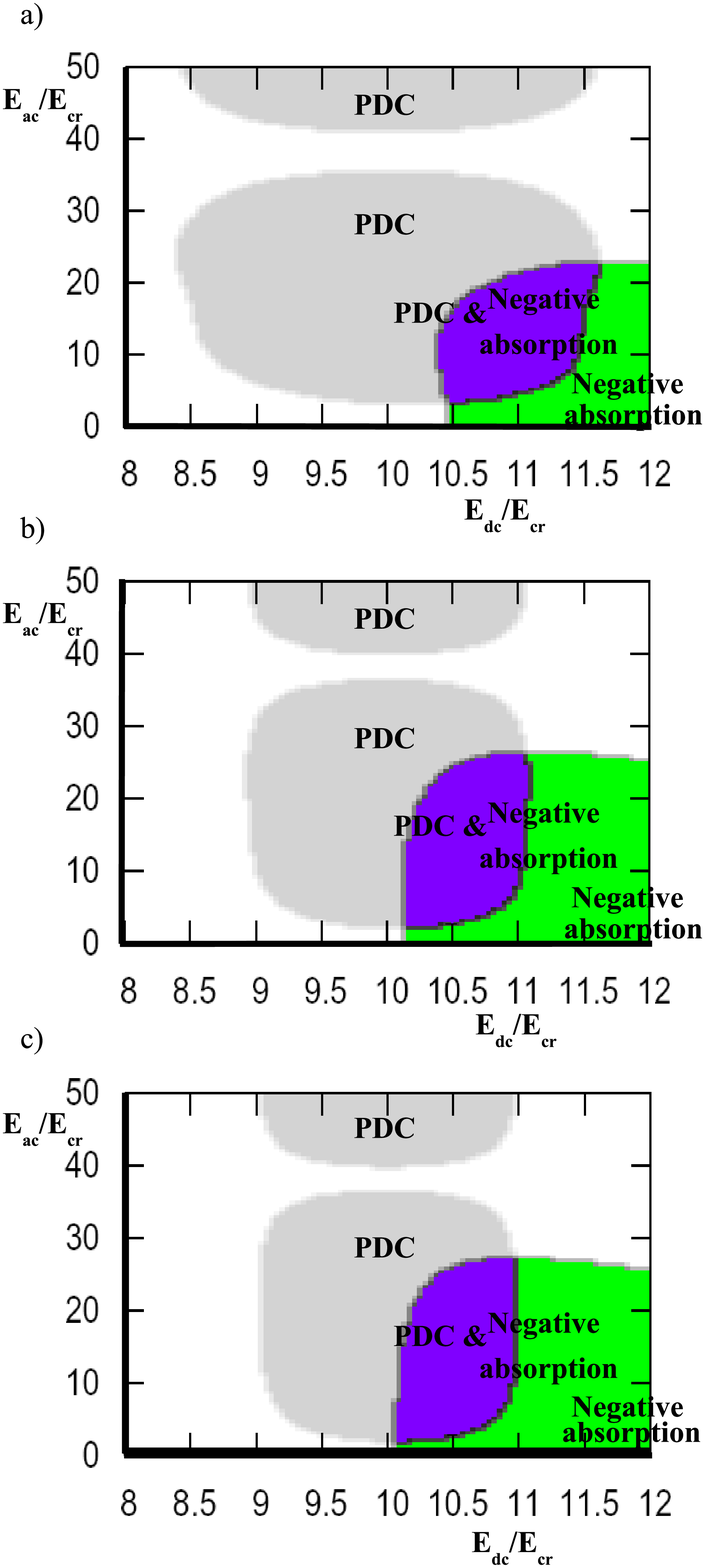}
  \caption{\label{fig:adjustabilities2} (Color online) Regions of $\sigma_0^\omega>0$
(grey) and $I_1^{\omega,\cos}<0$ (green) for $\omega
\tau_{\rm{eff}}=10$ with different values of $\nu$: (a) $\nu=0.1$,
(b) $\nu=0.4$ and (c) $\nu=1.0$. Overlapping of these regions
(violet) gives gain without instabilities.}
\end{figure}
% _____________________________________________________________________________
\par
On the other hand, even for a large enough ac field $E_{ac}$ the
absorption can stay negative for $\omega<\omega_B$, as can be
expected from the electron bunching mechanism \cite{Kroemer2}.
Figure~\ref{fig:acnfunktiona} shows the absorption $I_1^{\cos}$ and
the dc differential conductivity as functions of $E_{ac}$, when
$\omega \tau_{\rm{eff}}=10$ and $\omega_B\tau_{\rm{eff}}=10.8$. As
evident from this figure, there exists a well-defined range of
amplitudes $E_{ac}$ for which the positive differential conductivity
and the negative absorption occur simultaneously. When $\nu$
decreases the range of amplitudes $E_{ac}$ for negative absorption
shrinks and the range of amplitudes for PDC expands.
\par
The ranges of dc bias $E_{dc}$ and ac amplitude $E_{ac}$, supporting
PDC and gain at the first Shapiro-like step, are presented in
Fig.~\ref{fig:adjustabilities2} for three different values of $\nu$,
when $\omega \tau_{\rm{eff}}=10$. In this parameter plane, the area
of gain and simultaneous electric stability changes with a decrease
of $\nu$ (\textit{cf.} (a) and (c)). The change is very small as
long as $\nu \ge 0.4$.

\section{\label{Sec_time-scales}Characteristic time scales}

Until now we have considered only stationary transport properties.
Here we briefly examine and compare the time scales determining the
evolution of the fields in a superlattice. There are the following
characteristic times. Stationary VI characteristic is established
with the characteristic time $\tau$. (Single scattering time is
assumed for simplicity.) AC field defines the time scale
$2\pi/\omega$. Field inside an ideal (very high-$Q$) resonator is
growing with the characteristic time
%----------Eq 10-------------------------------------
\begin{equation}
\label{tau_g}
\tau_g=\frac{2 \epsilon \epsilon_0}{Re[\sigma(\omega)]}.
\end{equation}
%---------------------------------------------------
The characteristic time for domain formation is \cite{zakharov_60}
%----------Eq 11-------------------------------------
\begin{equation}
\label{tau_d}
\tau_d=\frac{\epsilon \epsilon_0}{\sigma_0(E_{dc})},
\end{equation}
%---------------------------------------------------
where $\sigma_0$ is the dc differential conductivity.
\par
We are working with a high-frequency fields satisfying
$\omega\tau>1$. The condition of LSA, additionally to the
requirement for a positive slope of time-averaged VI characteristic
(PDC), includes also the inequality
%----------Eqs 12-------------------------------------
\begin{equation}
\label{LSA2}
\omega\tau_d\gg 1.
\end{equation}
%---------------------------------------------------
It indicates a small (limited) charge accumulation during every THz
cycle. Condition (\ref{LSA2}) can be easily satisfied for typical
superlattices. Only in heavy doped superlattices with very wide
minibands, $\tau_d$ can be comparable to $\tau$ \cite{klap04}, and
therefore $\omega\tau_d=\omega\tau (\tau_d/\tau)$ can approach
unity.
\par
On the other hand, increasing the carrier density we are not only
increasing the gain but also reducing the time scale for the growth
of the field in the resonator (\ref{tau_g}). Moreover if
$\tau_g\ll\tau_d$, then the resonator-mode can reach the minimum
amplitude required for switching to local PDC before a domain would
be developed. Such a scenario can potentially solve the ``device
turn-on problem'' \cite{Kroemer} for Bloch oscillator. The ratio
$\tau_g/\tau_d$ depends only on $\omega$ and $\omega_B$, while
material parameters are not crucial. However, according to our
calculations $\tau_g$ and $\tau_d$ are close to each other for the
most interesting case $E_{dc}\gg E_{cr}$. This means that the
growing space charge domains and the growing ac field cannot be
handled separately.

% *****************************************************************************
% *****************************************************************************

\section{\label{sec:concl}Discussion and conclusion}

In summary, we have shown that THz gain in the conditions of
suppressed electric domains is possible in fairly large regions of
parameter space, which could allow to build devices which can be
used for generation and amplification of THz radiation. Large-signal
gain with suppressed domains is preserved with an introduction of
two different relaxation times for the electron velocity $\tau_v$
and energy $\tau_\varepsilon$. In particular, we demonstrated that
until $\tau_v/\tau_\varepsilon\gtrsim 0.5$, the difference from the
results obtained using a single $\tau$ is negligible. Note that
according to the experiments \cite{Schomburg}
$\tau_v/\tau_\varepsilon\gtrsim 0.5$ is a good assumption, although
it is not valid for all superlattices \cite{Hirakawa}.
Quantitatively, the magnitudes of all current components and thus
also the gain always decrease with decreasing ratio
$\tau_v/\tau_\varepsilon$.
\par
There remains, however, several important problems, which have not
been considered in the present work. The first one is the possible
influence of boundary conditions, which according to the
computational results of Rieder \cite{rieder} may turn out to be
crucial from the point of view of extended LSA mode. Hopefully, it
is still possible to control these boundary conditions at least to
some extent in which case the LSA mode can be made to work. Second,
we should mention that existence of LSA regime in superlattices at
THz frequencies is not completely proven even theoretically. The
existence of LSA operational mode is rather well established both
theoretically and experimentally in Gunn diodes at microwave
frequencies. However, still no experiments devoted to the LSA mode
in THz range are performed. From the theory side the consistent
derivation of LSA conditions has been done only within quasistatic
approximation  and thus it is not valid at THz frequencies.
Therefore, for those who like complete proofs, the existence of
extended LSA regime at THz frequencies still continues to be
interesting theoretical problem to study. Theoretical research in
these and related directions is in progress in Oulu. Of course,
experiments would also be very helpful in order to solve the
remaining important problems.
\par
Finally, in this work we have mainly focused on large-signal THz
gain in superlattices with suppressed space-charge instability. A
very important problem that still remains is how to get small-signal
gain without domains. One possible solution is to make use of
microwave pump to get gain at frequency multiplication in
superlattices \cite{Alekseev}.
\par
More full presentation of our research will be published
elsewhere.

% *****************************************************************************
% *****************************************************************************

\begin{acknowledgments}

We thank  Aleksey Shorokhov, Jukka Isoh\"{a}t\"{a}l\"{a} and Natalia
Alexeeva for the co-operation, as well as Gintaras Valu\v{s}is,
Stephan Winnerl and Alvydas Lisauskas for useful discussions. We are
grateful to Feo Kusmartsev and Piero Martinoli for constant
encouragement of this activity within EU programme. This research
was partially supported by Emil Aaltonen Foundation, Academy of
Finland and AQDJJ Programme of European Science Foundation.
\end{acknowledgments}

% *****************************************************************************
% *****************************************************************************

\end{document}